\documentclass[sigconf, balance=false, pbalance=true]{acmart}

\usepackage{booktabs}

\usepackage{amsmath,amssymb}
\usepackage{graphicx}
\usepackage{tikz}
\usetikzlibrary{positioning,arrows.meta}
\usepackage{pgfplots}
\pgfplotsset{compat=1.17}
\usepgfplotslibrary{groupplots}
\graphicspath{{../figures/}}

\renewcommand\footnotetextcopyrightpermission[1]{}
\acmConference{}{}{}

\makeatletter
\def\@titlefont{\Huge\rmfamily\bfseries}
\makeatother

\newcommand{\up}{\textsc{up}}
\newcommand{\down}{\textsc{down}}
\newcommand{\RES}{\mathrm{RES}}

\newcommand{\UNC}{\mathrm{UNC}}

\begin{document}
\microtypesetup{expansion=false}

\title{Ordinal Gates, Cardinal Bets: Matching LLM Confidence to the
Financial Decision Operator}

\author{Rayansh Singh}
\affiliation{%
  \institution{Michigan State University}
  \city{East Lansing}
  \country{USA}
}
\email{singhr26@msu.edu}

\author{Sara Rezaeimanesh}
\affiliation{%
  \institution{Michigan State University}
  \city{East Lansing}
  \country{USA}
}
\email{rezaeima@msu.edu}

\renewcommand{\shortauthors}{Rayansh Singh and Sara Rezaeimanesh}

\begin{abstract}
LLM confidence scores are not independently deployable objects: their decision value depends on the
downstream operator and exposure controller that consume them. Monotone recalibration cannot change a
coverage-matched rank-based gate, whereas position sizing consumes score magnitude, so changing a
confidence map can invalidate a scale fitted to the previous score distribution. We test this on
FactSet news for Nasdaq-100 equities, fitting maps and scales on 2021 and evaluating nine open-weight
LLMs out-of-sample on 2022--2023. Cross-applying raw and correctness maps with independently fitted
scales shows that the two components are not portable alone: scale transfer reduces certainty-equivalent return (CER) in $8/9$
models and produces large risk-target errors. Matching each map with its fitted scale improves
ensemble CER by $9.2$ percentage points per year under frozen-scale control ($p<0.001$),
and the effect remains significant when the single largest-contributing model is excluded
($+5.5$pp/yr), so it is not driven by one case. Under an identical adaptive-volatility
controller, however, the incremental effect falls to $+1.6$pp/yr, with a significant
controller interaction. Annual walk-forward effects are smaller, although map--scale interaction remains positive in every
fold. Confidence transformations should therefore be evaluated jointly with the downstream
controllers that consume them.
\end{abstract}


\maketitle

\section{Introduction}

One confidence score can determine both whether an AI system acts and how strongly it
acts~\cite{chow1970,geifman2017}: an instruction-tuned LLM asked to classify a financial news
headline as predicting an \up{} or \down{} next-day return returns a confidence that can either
gate a position (the classical reject option) or size it. We make this distinction precise
through the \emph{decision operator}, the map from a confidence signal to an action, and build our
central claim around it: a confidence score's decision value is operator-relative---ordinal
operators consume only its ranking, while cardinal operators consume its magnitude and therefore
depend on the controller paired with it.

Prior financial-LLM work has examined directional return signals and documented confidence
miscalibration~\cite{lopezlira2023,chen2024}, while calibration research largely evaluates
statistical reliability or threshold-based selection as ends in themselves~\cite{guo2017,naeini2015}.
Prior work has not directly isolated which property of the score the downstream decision actually
requires.
Contemporaneous work makes the distinction concrete: \citet{iacovides2025findpo} temperature-scale
a saturated sentiment score before ranking assets into fixed portfolio quantiles (Section~\ref{sec:related}).
We build on the fact that monotone calibration cannot reorder
predictions~\cite{rabanser2025selective} and extend the observation to the less-studied
cardinal case of position sizing.

\emph{Ordinal operators} (coverage-matched thresholding, top-$k$ selection, abstention, routing)
consume only rank: they are invariant to strictly increasing transformations and therefore require
discrimination rather than calibrated magnitude. \emph{Cardinal operators} (sizing, allocation,
portfolio weighting) consume magnitude, so their scores must align with the economic target the
operator requires---ideally a conditional signed payoff relative to risk, rather than correctness
alone (Section~\ref{sec:cardinal}). A mismatched confidence map and controller can produce a material
risk-budget error, in the clearest case running a raw system at up to $2.8\times$ its intended risk
out of sample (Section~\ref{sec:ninemodel}). This is a model-governance problem: replacing the LLM, elicitation
method, or calibration map can silently invalidate an exposure controller fitted to the previous
score distribution.

We test this distinction on FactSet professional-newswire headlines for Nasdaq-100 constituents
(2021--2023), fitting every map, threshold, and exposure scale on 2021 and evaluating forward on
2022--2023 under look-ahead-free execution. The ordinal claim is tested across five
confidence-extraction channels on two open-weight LLMs (Qwen2.5-7B, Gemma-2-9B); the cardinal claim
is tested with verbalized confidence across a nine-model grid spanning 7--32B parameters and five
model families, comparing each raw-confidence system against a matched system pairing a
holding-period correctness map with its own frozen-scale controller
(Section~\ref{sec:theory} details the four-system design). The ordinal analysis uses next-session
classification correctness, whereas the sizing analysis fits a holding-period correctness map
aligned with the implemented five-session position (Section~\ref{sec:data-splits}).

Empirically, ordinal recalibration repairs reliability without changing coverage-matched rankings,
while the tested confidence channels have little resolution for selective prediction
(Section~\ref{sec:main}). For cardinal sizing, cross-applying confidence maps and independently
fitted exposure scales produces large risk distortions, whereas pairing each map with its own fitted
scale improves ensemble CER by $9.2$pp/yr under frozen-scale control---an incremental gain that
largely dissolves under an identical adaptive-volatility controller, exposing the map--scale coupling
that the results section quantifies (Sections~\ref{sec:sizing-exp}--\ref{sec:walkforward}).

We make two contributions: (1) we formalize the consequences of the ordinal--cardinal distinction for
LLM confidence in financial decisions (Section~\ref{sec:theory}); and
(2) we show empirically that confidence maps and fitted exposure controls are coupled rather than
independently portable. The latter coupling---not the headline matched-system CER gain---is the
paper's general design implication.

Our aim is to study how confidence transformations interact with the financial operator and
exposure-control procedure that consume them. We do not treat correctness as a universal sizing target
or claim regime invariance.

\section{Related Work}\label{sec:related}

\paragraph{Confidence calibration and selective decisions.} Calibration research measures
reliability with ECE~\cite{naeini2015} and Brier score~\cite{brier1950} and repairs it with
temperature scaling~\cite{guo2017}, treating reliability as an end in itself rather than a
selective-classification signal. Decision-relative calibration is itself established---%
\citet{zhao2021decision} formalize calibration against downstream decision-makers, and
\citet{bas2026} evaluate LLM confidence by downstream utility---so our claim is narrower: two
financial operators consume different properties of the same signal (rank vs.\ magnitude), making
calibration and exposure control non-portable modules. Verbalized confidence is
a studied elicitation channel~\cite{tian2023,xiong2024,kadavath2022}, and we make its classical
distinction from \emph{refinement}~\cite{degroot1983} decision-relative. The reject option dates to
Chow~\cite{chow1970}, and deep selective classification to Geifman and El-Yaniv~\cite{geifman2017}.
Post-hoc recalibration routinely improves threshold-based rejection~\cite{joy2023adaptive,guo2017}.
But \citet{rabanser2025selective} show monotone calibration cannot reorder predictions, so it cannot
close the selective-classification gap; our rank-invariance result in Section~\ref{sec:theory} is the
formal basis.

\paragraph{Financial LLMs and exposure sizing.} Contemporaneous work on
FinDPO~\cite{iacovides2025findpo} illustrates the same gap concretely: it extracts a softmax
sentiment score from class-token logits, observes the same near-$0$/$1$ overconfidence saturation we
document (Section~\ref{sec:main}), and applies temperature scaling to it---yet its portfolio ranks
assets by that score and equal-weights a fixed top/bottom quantile, an ordinal operator that
Section~\ref{sec:ordinal} shows is invariant to exactly this fix. Other recent work examines
adjacent failure modes---prompting-strategy accuracy~\citep{vamvourellis2025reasoning}, positional
bias~\citep{dimino2025positional}, and sector\slash size\slash momentum bias~\citep{lee2025yourai}---without
asking whether a score's magnitude is safe to consume for sizing versus ranking, and we test both
sides of that distinction---the rank operator and its cardinal utility counterpart
(Section~\ref{sec:cardinal})---within one LLM confidence system. Mapping a forecast into a position size is
classical: mean-variance choice~\cite{markowitz1952}, Kelly sizing~\cite{kelly1956}, and the
``fundamental law''~\cite{grinoldkahn2000} convert a signal into exposure via its first two moments;
volatility targeting~\cite{moreiramuir2017} and parameter-uncertainty~\cite{demiguel2009} are handled
empirically by our frozen-scale controller, fit \emph{after} calibration. For look-ahead bias we
adopt Glasserman and Lin's~\cite{glasserman2023} named-vs-anonymized protocol under forward execution
(Section~\ref{sec:data}).

\section{Theory}\label{sec:theory}

\paragraph{Setup.} A classifier emits direction $D(X)\in\{-1,+1\}$ and confidence
$C(X)\in[\tfrac12,1]$ separately. Define correctness $Z=\mathbf 1[D(X)=Y]$ and the
correctness-calibration function $\mu(u):=P(Z=1\mid C=u)$~\cite{murphy1973,degroot1983}.

\paragraph{Terminology.} We fix five terms, chained by how a confidence score becomes an exposure:
\emph{map} $g$ (raw $g_{\rm raw}=\mathrm{id}$, correctness $g_{\rm corr}=\hat\mu$) $\to$
\emph{shape} $h$ ($\phi$ when scale-free) $\to$ \emph{scale} $\kappa$ (a parameter, not a controller)
$\to$ \emph{controller} (shape $+$ scale $+$ per-name caps $+$ gross normalization; e.g.\ the
frozen-scale and adaptive-volatility controllers of Section~\ref{sec:ninemodel}) $\to$
\emph{system} (one map $+$ one controller). The four-cell experiment tests whether maps and fitted
scales are compatible; we use \emph{map--controller compatibility} for the broader deployment claim.

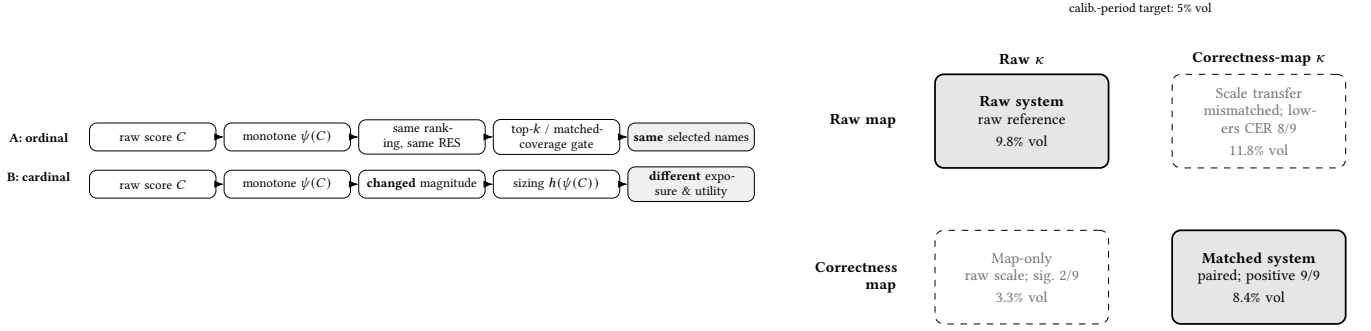
\begin{figure*}[t]
\centering
\begin{minipage}[c]{0.56\textwidth}
\centering
\resizebox{\linewidth}{!}{%
\begin{tikzpicture}[
  node distance=4pt,
  every node/.style={font=\footnotesize},
  box/.style={draw, rounded corners, align=center, minimum height=15pt, text width=64pt, inner sep=2pt},
  tag/.style={font=\footnotesize\bfseries},
  arr/.style={-{Latex[length=5pt]}}
]
\node[tag] (Atag) {A: ordinal};
\node[box, right=8pt of Atag] (Ac) {raw score $C$};
\node[box, right=of Ac] (Apsi) {monotone $\psi(C)$};
\node[box, right=of Apsi] (Arank) {same ranking, same $\mathrm{RES}$};
\node[box, right=of Arank] (Atopk) {top-$k$ / matched-coverage gate};
\node[box, right=of Atopk, fill=gray!12] (Asame) {\textbf{same} selected names};
\draw[arr] (Ac)--(Apsi); \draw[arr] (Apsi)--(Arank);
\draw[arr] (Arank)--(Atopk); \draw[arr] (Atopk)--(Asame);

\node[tag, below=10pt of Atag] (Btag) {B: cardinal};
\node[box, below=10pt of Ac] (Bc) {raw score $C$};
\node[box, right=of Bc] (Bpsi) {monotone $\psi(C)$};
\node[box, right=of Bpsi] (Bmag) {\textbf{changed} magnitude};
\node[box, right=of Bmag] (Bsize) {sizing $h(\psi(C))$};
\node[box, right=of Bsize, fill=gray!12] (Bdiff) {\textbf{different} exposure \& utility};
\draw[arr] (Bc)--(Bpsi); \draw[arr] (Bpsi)--(Bmag);
\draw[arr] (Bmag)--(Bsize); \draw[arr] (Bsize)--(Bdiff);
\end{tikzpicture}%
}
\end{minipage}%
\hfill
\begin{minipage}[c]{0.40\textwidth}
\centering
\resizebox{\linewidth}{!}{%
\begin{tikzpicture}[
  every node/.style={font=\footnotesize},
  cellprimary/.style={draw, thick, rounded corners, align=center, text width=2.5cm,
    minimum height=1.5cm, inner sep=4pt, fill=gray!20},
  celloff/.style={draw, dashed, rounded corners, align=center, text width=2.5cm,
    minimum height=1.5cm, inner sep=4pt, text=black!55},
  hdr/.style={font=\footnotesize\bfseries}
]
\node[hdr, align=center, text width=2.5cm] (colraw) at (1.4,2.5) {Raw $\kappa$};
\node[hdr, align=center, text width=2.5cm] (colcal) at (5.2,2.5) {Correctness-map $\kappa$};
\node[hdr, anchor=east, align=right, text width=1.3cm] (rowraw) at (-0.5,1.5) {Raw map};
\node[hdr, anchor=east, align=right, text width=1.3cm] (rowcal) at (-0.5,-1.0) {Correctness map};
\node[font=\scriptsize, align=center] at (3.3,3.3) {calib.-period target: 5\% vol};
\node[cellprimary] (rr) at (1.4,1.5) {\textbf{Raw system}\\raw reference\\[2pt]9.8\% vol};
\node[celloff] (rc) at (5.2,1.5) {Scale transfer\\mismatched; lowers CER 8/9\\[2pt]11.8\% vol};
\node[celloff] (cr) at (1.4,-1.0) {Map-only\\raw scale; sig.\ 2/9\\[2pt]3.3\% vol};
\node[cellprimary] (cc) at (5.2,-1.0) {\textbf{Matched system}\\paired; positive 9/9\\[2pt]8.4\% vol};
\end{tikzpicture}%
}
\end{minipage}
\caption{Left: the same monotone transformation preserves ranking (ordinal, \S\ref{sec:ordinal})
but changes magnitude (cardinal, \S\ref{sec:cardinal}). Right: the four-system map--scale design
(\S\ref{sec:cardinal}); the shaded diagonal cells (raw system, matched system) are the primary
comparison, and the dashed off-diagonal cells diagnose whether maps and scales are portable across
signal distributions. Volatility values are mean per-model realized test-period volatility
(distinct from the equal-weight ensemble-portfolio volatility of Table~\ref{tab:matched}).}
\label{fig:operators}
\end{figure*}

\subsection{Ordinal operators: a resolution bound on gate accuracy}\label{sec:ordinal}
\label{prop}%

A \emph{coverage gate} retains the top-$\varphi$ fraction by score. Its accuracy gain
$\Delta_{\mathrm{acc}}(\tau)=P(Z=1\mid C\ge\tau)-P(Z=1)$ satisfies
\begin{equation}\label{eq:covidentity}
\Delta_{\mathrm{acc}}(\tau)=\varphi^{-1}\operatorname{Cov}(\mu(C),\mathbf 1[C\ge\tau]),
\end{equation}
and by Cauchy--Schwarz, writing $\mathrm{RES}(C)=\operatorname{Var}(\mu(C))$ for Brier
resolution~\cite{murphy1973},
\begin{equation}\label{eq:cs}
|\Delta_{\mathrm{acc}}(\tau)|\le\sqrt{\mathrm{RES}(C)\,\tfrac{1-\varphi}{\varphi}}.
\end{equation}
For every strictly increasing $\psi$, the threshold family $\{C\ge\tau\}_\tau$ bijects onto
$\{\psi(C)\ge t\}_t$, so the accuracy--coverage frontier, every coverage-matched portfolio, and
$\mathrm{RES}$ are \emph{invariant} to $\psi$. A fixed numerical threshold (confidence above $0.70$,
say) is magnitude-sensitive, not rank-only, so it falls outside this invariance result---only
coverage-matched thresholds, which adapt to the score distribution, are covered. Temperature and
Platt scaling are strictly increasing, so they can improve reliability but cannot change score
order, $\mathrm{RES}$, or a coverage-matched gate (slopes verified positive for all 18
configurations; Section~\ref{sec:main}).
Isotonic regression is only \emph{nondecreasing}: it may introduce plateaus, so this exact bijection
claim applies to strictly increasing $\psi$; in our empirical isotonic check, ties within a plateau
are broken deterministically by the pre-calibration raw score, which we verify preserves the
coverage-matched selected set exactly (0 discrepancies across all 18 ordinal configurations and 20
coverage levels from 5\% to 100\%).

\subsection{Cardinal operators: map-specific scales and magnitude sensitivity}\label{sec:cardinal}
\label{prop:sizing}%

The following oracle benchmark isolates why a magnitude-consuming operator can benefit from a
better-aligned score even when rank is unchanged; it is not the system implemented in
Section~\ref{sec:sizing-exp}, and the gap between them is made explicit below. A confidence-sized
position earns $D(X)h(C(X))R$. Write $Q=D(X)R$, $m(u)=E[Q\mid C=u]$,
$v(u)=\operatorname{Var}(Q\mid C=u)$. Maximizing one-period conditional quadratic utility
$E[h(C)Q]-\tfrac{\lambda}{2}E[v(C)h(C)^2]$ ($\lambda>0$) over $h$ gives the oracle sizing rule
\begin{equation}\label{eq:hstar}
h^\star(u) = \frac{m(u)}{\lambda v(u)} \ \propto\ \frac{E[Q\mid C=u]}{\operatorname{Var}(Q\mid C=u)}.
\end{equation}
\emph{The cardinal target is therefore conditional signed payoff relative to risk, not correctness
probability alone.}

Equation~\eqref{eq:hstar} identifies conditional signed payoff relative to risk as the oracle target.
Our primary system instead uses correctness probability as a lower-variance proxy and restricts sizing
to a map-specific shape multiplied by a scalar scale. We therefore test whether changing the confidence
map requires refitting that scale. Section~\ref{sec:mechanism} separately evaluates direct signed-payoff
estimation.

The implemented system does not fit $h$ freely: it fixes a map-specific shape $\phi(C)\in[0,1]$ and a
single scalar $\kappa\ge0$, $h(C)=\kappa\,\phi(C)$ (Section~\ref{sec:data-portfolio},
eq.~\eqref{eq:signal}). Restricting the objective above to this family, writing $Y_\phi:=\phi(C)Q$ and
$v_\phi:=E[v(C)\phi(C)^2]$, gives $U_\phi(\kappa)=\kappa E[Y_\phi]-\tfrac\lambda2\kappa^2 v_\phi$; when
$E[Y_\phi]>0$ the restricted optimum is $\kappa_\phi^\star=E[Y_\phi]/(\lambda v_\phi)$.

Cross-applying another shape's optimum incurs transfer regret
$U_\phi(\kappa_\phi^\star)-U_\phi(\kappa_{\phi'}^\star)=
\tfrac\lambda2 v_\phi(\kappa_{\phi'}^\star-\kappa_\phi^\star)^2\ge0$ on the distribution defining
$\kappa_\phi^\star$, so \textbf{the exposure scale is shape-specific}---what the four-system design
below examines under estimated scales, caps, costs, and temporal shift.

The empirical volatility-targeting rule
$\kappa_\phi^{\mathrm{vol}}=\sigma_{\mathrm{target}}/\operatorname{SD}(Y_\phi)$ is the analogous
map-specific construction (Section~\ref{sec:ninemodel}).

\paragraph{Scale equivariance.} Uniform positive rescaling cannot explain the matched effect because
refitting the volatility scale exactly offsets it before caps bind; improvement therefore requires
nonlinear reshaping, support changes, cap interactions, or different out-of-sample behavior. Thus any
matched-system improvement must arise from changes in relative magnitudes or in the active-position
support, rather than from a trivial global rescaling of confidence.

\paragraph{The four-system design.} We distinguish the confidence transformation $g_m(C)$ from the
sizing function it feeds: $g_{\rm raw}(C)=C$, $g_{\rm corr}(C)=\hat\mu(C)$, and
$h_m(C)=\max(0,2g_m(C)-1)$ for either map (eq.~\eqref{eq:signal}). Crossing the two confidence maps
($g_{\rm raw}$, $g_{\rm corr}$) with two independently fitted scales $\kappa$ (raw; correctness-map)
yields four systems: \emph{raw} (raw map, raw scale), \emph{map-only} (correctness map, raw scale),
\emph{scale-transfer} (raw map, correctness-map scale), and \emph{matched} (correctness map,
correctness-map scale). The two diagonal systems are valid map--scale pairings; the two off-diagonal
systems diagnose whether a map and a scale are portable across signal distributions. The map-only
cell, which holds the exposure scale fixed while the map changes, is our primary shared-scale
surrogate (Figure~\ref{fig:operators}).

Writing $J(g,\kappa)$ for realized test-period CER under the actual caps, costs, and normalization of
Section~\ref{sec:data-portfolio}, and $J_{rr},J_{cr},J_{rc},J_{cc}$ for the four systems, the matched
effect decomposes exactly---with no distributional assumption---into map, scale, and
map$\times$scale-interaction terms:
\begin{equation}\label{eq:decomp}
J_{cc}-J_{rr}=(J_{cr}-J_{rr})+(J_{rc}-J_{rr})+(J_{cc}-J_{cr}-J_{rc}+J_{rr}).
\end{equation}
Section~\ref{sec:ninemodel} estimates each component---map replacement, scale transfer, and their
interaction---directly under the implemented caps, costs, and temporal split.


\section{Data and Evaluation Protocol}\label{sec:data}

\subsection{Data, splits, and labels}\label{sec:data-splits}

We use FactSet professional-newswire headlines for Nasdaq-100 constituents (2021--2023). The
cleaning pipeline restricts to US-equity, English-language wire items, strips source prefixes, 
and drops routine ownership/sector calendar notices. The unit of analysis is the 
\emph{(ticker, trading-day)}. All headlines for a ticker-day are aggregated into
one bundle (mean $4.17$/bundle, median $2$). Calibration is 2021 ($15{,}034$ ticker-days) and
evaluation is 2022--2023 ($31{,}888$ ticker-days, $505$ test-period trading sessions). Every
recalibration parameter, temperature, frozen-scale controller, and gate threshold $\theta^\star$ is fit
only on 2021. We run \emph{named} and \emph{anonymized} conditions (company names/tickers masked with
\texttt{[COMPANY]} \emph{before} the LLM sees the text) as separate configurations throughout.

FactSet does not provide granular intraday timestamps, so a ticker-day bundle dated $D$ is assigned to the first trading session
strictly after $D$--session $T$--to eliminate same-session execution. The ordinal experiment
(Section~\ref{sec:main}) evaluates whether the predicted direction matches that session's own
open-to-close return sign. The cardinal experiment (Section~\ref{sec:ninemodel}) sizes a five-session
position, so its downstream-success map instead estimates whether that same predicted direction
matches the five-session holding-period return sign. This is a decision-aligned success target, not a
recalibration of the original one-session prediction. The exact
ordinal invariance result is horizon-independent; the one-session gate is the classification-aligned
illustration, whereas the five-session target corresponds to the implemented holding-period decision
(direct payoff estimation targets this quantity directly; Section~\ref{sec:mechanism} compares it to
correctness calibration). Five sessions was fixed a priori as the primary weekly holding cadence. A descriptive twelve-horizon
sensitivity sweep places $h{=}5$ within a broad stable region: out-of-sample matched effects remain
positive through $h{=}8$ and attenuate at longer horizons; no multiplicity-adjusted inference is
claimed across horizons.

\paragraph{Primary sample and temporal extension.} The primary nine-model evaluation
(Section~\ref{sec:ninemodel}) uses the 2021--2023 sample above. The temporal extension
(Section~\ref{sec:walkforward}) evaluates a five-model subset\footnote{Qwen2.5-7B, Gemma-2-9B, Phi-4,
Mistral-24B, and Gemma3-27B.} spanning model families and parameter scales on $12{,}911$ additional
FactSet 2024--2025 ticker-days using four annual walk-forward folds. Each fold fits the confidence
map and both exposure scales on year $t$ and evaluates them frozen on year $t{+}1$, holding horizon,
costs, caps, universe, and $\gamma$ fixed. Signals whose five-session hold crosses the
calibration-year boundary are dropped, not truncated, and bootstrap blocks are sampled within fold
boundaries. Each test year is forward-held-out relative to its calibration fit, though not
necessarily relative to model pretraining.

\subsection{Models and confidence signals}\label{sec:data-models}

\paragraph{Ordinal grid.} Qwen2.5-7B and Gemma-2-9B across five confidence-extraction channels,
chosen because they preserve different statistical properties of the same model: some
rank well but are miscalibrated, others the reverse, making the operator-relative claim
testable rather than definitionally true of one extraction method. \emph{Verbalized} (regex-parsed
``Direction/Confidence,'' parse-failure $2.6$--$3.5\%$, recorded as missing and excluded from that
ticker-day's active signal set, never imputed),
\emph{two-token logit} ($\Delta z=\log p_\up-\log p_\down$, predicted-class confidence
$\sigma(|\Delta z|)$), \emph{logit-full} (full-vocabulary \up/\down{} probability mass),
\emph{logit-ent} (normalized negentropy of the first-token distribution), and \emph{P(True)}
(Qwen2.5 only~\cite{kadavath2022}); a $K{=}5$ temperature-sampled self-consistency
baseline~\citep{wang2022self} is reported separately as an additional baseline, outside this family.
The four channels common to both models give 16 configurations, and Qwen-only P(True) in both
conditions adds two, for \textbf{18 model$\times$condition configurations} across the five primary
channels---the accuracy-tuned family tested jointly in Section~\ref{sec:main}.

\paragraph{Cardinal grid.} Nine unmodified HuggingFace open-weight
checkpoints\footnote{Models: Qwen2.5-\{7B,14B,32B\}, Gemma-2-9B, Gemma-3-27B, Mistral-\{7B,24B\},
Phi-4, and FinLLaMa-8B.} spanning 7--32B parameters across five model families (Qwen, Gemma, Mistral,
Phi-4, FinLLaMa), using verbalized confidence, the only channel consistently available across all
checkpoints. The 2025
extension corpus (Section~\ref{sec:walkforward}) postdates the two primary models' release, ruling
out pretraining-data inclusion for them. Pretraining-cutoff audit status for the other seven is a
limitation (Section~\ref{sec:limits}).

\paragraph{Implementation.} Each bundle is rendered as a single user-role chat-template turn (no
system prompt), truncated to 512 input tokens. Multi-headline bundles are pipe-joined. Verbalized
decoding is greedy. The logit channel is a single forward
pass with no generation. Models $\ge24$B parameters run int8-quantized; all others run unquantized
float16. All random seeds are fixed to 42. The verbalized channel asks the model to predict UP or
DOWN and report an integer confidence from 0 to 100 (50 = completely uncertain).

\subsection{Portfolio construction and inference}\label{sec:data-portfolio}

Each ticker-day bundle produces one direction $D_{i,t}\in\{-1,+1\}$ and confidence $C_{i,t}$ per
model$\times$condition. For map $m\in\{\text{raw},\text{correctness}\}$, the signed magnitude, capped
position, and daily portfolio return are
\small
\begin{align}
s_{i,t}^{(m)} &= D_{i,t}\,\max\!\big(0,\,2\,g_m(C_{i,t})-1\big), \label{eq:signal}\\
W_{i,t}^{(m)} &= \tfrac{1}{5}\!\sum_{\ell=0}^{4}\kappa_m s_{i,t-\ell}^{(m)}, \label{eq:aggexposure}\\
w_{i,t}^{(m)} &= \mathrm{clip}\!\big(W_{i,t}^{(m)},\pm0.10\big)\cdot
  \min\!\Big(1,\tfrac{1.0}{\sum_j|\mathrm{clip}(W_{j,t}^{(m)},\pm0.10)|}\Big), \label{eq:weight}\\
r_t^{(m)} &= \sum_i \frac{w_{i,t}^{(m)}}{W_{i,t}^{(m)}}
  \sum_{\ell=0}^{4}\frac{\kappa_m s_{i,t-\ell}^{(m)}}{5}\,R_{i,t}^{(\ell)}
  - c\,\mathrm{Turnover}_t^{(m)}, \label{eq:portreturn}
\end{align}
\normalsize
where $g_m$ is the identity for the raw-confidence map, and the correctness map uses the fitted
holding-period correctness function $\hat\mu(\cdot)$, which we estimate by isotonic calibration.
Isotonic regression is the primary calibrator; Beta and Platt regression are
robustness specifications, fit and reported identically but never substituted for isotonic in a
primary result. Per-name and gross-normalization caps apply to the aggregate outstanding holding $W_{i,t}^{(m)}$
summed across all five overlapping entry cohorts, not to each cohort individually: $W_{i,t}^{(m)}$ is
$\kappa_m$-scaled, clipped to a $10\%$ per-name cap, then rescaled down (never up) to a $100\%$ gross
cap, and the same cap-induced scale factor $w_{i,t}^{(m)}/W_{i,t}^{(m)}$ multiplies every cohort leg,
so daily gross exposure stays at or below $1$. Here $R_{i,t}^{(0)}$/$R_{i,t}^{(\ell>0)}$ are the
entry-session open-to-close leg and subsequent close-to-close legs of the five-session hold, entered
at $T$'s own open (the classification-correctness label's session and first executable price). Each
position change is charged $c=5$bps one-way, no overnight borrow charge. The candidate universe is the Nasdaq-100 constituent list as of January 2021, yielding $88$ tickers;
$7$ names lack vendor price coverage, leaving $81$, of which $78$ have $\ge1$ 2021 headline. The universe is frozen to
$61$ tickers with $\ge30$ ticker-days in 2021 calibration, applied unchanged to 2022--2023. Nasdaq-100
additions after 2021 are excluded to keep the universe ex-ante, while four acquired/delisted
constituents lacking vendor prices are absent. All prices use FactSet split and dividend adjustments.

$\kappa_m$ is selected on 2021 returns to target $5\%$ annualized pre-cost volatility ($100\%$
gross, $10\%$ per-name cap). All maps, thresholds, and scales are frozen before evaluation. CER with
$\gamma=3$ ($\mathrm{CER}_\gamma=252(\bar r-\gamma\operatorname{Var}(r)/2)$) is the primary economic
outcome. The primary test uses a joint circular block bootstrap of the model-ensemble matched effect
with $5{,}000$ replicates and block length 5. Block lengths 10 and 20 are sensitivity checks, and all
tests are two-sided. The estimand is this fixed ensemble on the observed path, not a population of
LLMs or regimes. Model-level $p$-values are Holm-corrected across the nine-model family, and the
fixed-threshold gate (Section~\ref{sec:main}) is tested jointly across all 18 configurations with
Romano--Wolf~\cite{romano2005} stepdown control. Across the nine-model grid, the matched system
lowers mean per-model realized volatility (condition-averaged, 9 models) from $9.9\%$ to $8.3\%$ and
average daily active positions from $18.1$ to $14.5$; the equal-weight model-ensemble portfolio's
volatility falls from $8.8\%$ to $7.8\%$ (Table~\ref{tab:matched}). Realized portfolio constraints
were never binding: mean daily gross exposure was $85.2\%$, with a $95\%$th percentile of $98.1\%$
and maximum of $99.8\%$; per-name caps were never exceeded.

\paragraph{Evidentiary status.} The pre-specified primary economic estimand is the
2021-fit/2022--2023 matched-versus-raw ensemble effect under frozen-scale control; alternative
calibrators, model exclusions, placebos, controller choice, costs, and annual transfer are
confirmatory robustness checks, while the fixed-2024H1, diagnostic-correlation, label-noise, and
subgroup analyses are exploratory.

\section{Ordinal Decisions}\label{sec:main}

\paragraph{Calibration without decision change.} Taking $\mathrm{softmax}([z_{\up},\allowbreak z_{\down}])$ on two vocabulary
logits and using the max as confidence saturates: LLM vocabulary logits routinely differ by $20+$
nats, collapsing confidence to $\approx0.97$--$0.98$ and ECE to $\approx0.47$ (both models, both
conditions). Fitting one temperature $T$ per model on the calib split repairs this: logit ECE falls
to $0.03$--$0.04$. Because softmax and $\sigma(\Delta z/T)$
are both strictly increasing in $\Delta z$, they induce the identical ranking
(Section~\ref{sec:ordinal}): the coverage-matched gate is exactly unchanged even though ECE moves by
an order of magnitude.

\paragraph{Low resolution limits selective gains.} Resolution---the degree to which confidence
separates observations with different conditional success rates---is the quantity a coverage gate
needs. The Murphy decomposition makes the binding constraint
precise: with $\UNC=\bar p(1-\bar p)\approx0.249$, resolution is small everywhere,
$\RES/\UNC$ ranging $0.07\%$--$0.4\%$ across all 18 configurations, and the base classifier itself is
weak (1-session accuracy $50.3$--$51.8\%$, at or below the $53.8\%$ base rate). Bound~\eqref{eq:cs}
caps the \emph{accuracy} gain any coverage gate can extract from resolution this low: at the $10\%$
coverage floor used below, roughly $4$--$9$ percentage points even at best-resolution. The same
conclusion holds across the five primary channels (verbalized, three logit variants, and P(True));
none provides enough resolution for dependable selection. A $K{=}5$ temperature-sampled
self-consistency baseline~\citep{wang2022self}, reported separately, adds none---its outputs are
unanimous on $96$--$98\%$ of ticker-days.

\paragraph{Fixed-threshold gating yields no reliable economic lift.} Fixed numerical thresholds fall
outside the rank-invariance result above because they consume score magnitude. Thresholds fitted on
2021 either for accuracy or CER nevertheless yield no family-wise significant 2022--2023
improvement: 0/18 accuracy-tuned configurations survive Romano--Wolf correction, and 0/18 CER-tuned
configurations survive Holm correction despite 15/18 positive point estimates. Thus, recalibration
cannot change coverage-matched selection, while the available score rankings are too weak to support
reliable fixed-threshold gains.

\section{Cardinal Decisions}\label{sec:sizing-exp}\label{sec:ninemodel}

\subsection{Maps and scales are not independently portable}\label{sec:robust}\label{sec:robust-compact}

\paragraph{Portability diagnosis.} Maps and exposure scales perform poorly when transferred
independently.

Cross-applying each map with each fitted scale (Figure~\ref{fig:operators}, right) shows neither is
portable alone: map replacement at the raw scale is Holm-significant in only $2/9$ models, while the
correctness-map scale applied to the raw map reduces CER in $8/9$. The resulting model-ensemble
map--scale interaction (eq.~\eqref{eq:decomp}'s last term) is $+17.1$pp/yr; we interpret this
interaction descriptively rather than causally. Consistent with map-specific scaling, the
correctness-map scale realizes $11.8\%$ mean per-model test volatility when transferred to the raw
map, versus $8.4\%$ with its intended correctness map. Conversely, the raw-map scale realizes $9.8\%$
with its intended raw map but only $3.3\%$ when transferred to the correctness map. Holding the scale
fixed while changing the confidence map therefore shifts realized risk sharply in either direction;
matching reduces transfer distortion but does not eliminate fit-to-test risk drift. The next
two subsections show what pairing each map with its own fitted scale buys back, and how much of that
recovery is magnitude reallocation versus filtering. Applying the correctness map to a
coverage-matched gate, meanwhile, leaves the portfolio unchanged, as predicted by
Section~\ref{sec:ordinal}.

\subsection{Matched map--scale pairing improves the frozen system}\label{sec:primary-matched}

\paragraph{Matched pairing reduces transfer distortion.} Pairing each holding-period correctness map
with its own fitted
frozen-scale controller removes the cross-map scale mismatch diagnosed in
Section~\ref{sec:robust-compact} and improves ensemble CER by $9.2$pp/yr.

We compare each \emph{raw} system (raw map, own fitted $\kappa$) with the corresponding
\emph{matched} system (correctness map, own fitted $\kappa$), condition-averaging named and anonymized
inputs; the off-diagonal systems of Section~\ref{sec:robust-compact} serve as portability diagnostics.
The map-specific scalar benchmark of Section~\ref{sec:cardinal} predicts that scales are generally not
portable across maps, but whether the correctness diagonal outperforms the raw diagonal under estimated
scales, caps, costs, and temporal shift is an empirical question we estimate directly across the nine
models (Section~\ref{sec:data-models}).

\begin{table*}[t]
\caption{Raw-versus-matched comparison, $\gamma=3$, pp/yr CER, 2022--2023 test,
condition-averaged across named and anonymized inputs per model (joint circular block bootstrap,
same draw across raw/matched). $^{**}$: Holm-significant at $\alpha=0.05$ across the 9-model
family. The model-ensemble row is bootstrapped as one equal-weight portfolio with common date
blocks, not nine independent $p$-values.}
\label{tab:matched}
\small
\setlength{\tabcolsep}{6pt}
\begin{tabular}{@{}lrrrrc@{}}
\toprule
Model & Raw CER & Matched CER & $\Delta$CER & $95\%$ CI & Vol.\ raw$\to$matched \\
\midrule
Qwen2.5-7B  & $-25.5$ & $14.2$ & $\mathbf{+39.7}^{**}$ & $[24.5,56.1]$ & $12.3\%\to7.3\%$ \\
Gemma-2-9B  & $11.5$  & $16.0$ & $+4.5$  & $[0.1,9.2]$    & $9.5\%\to8.7\%$ \\
Mistral-7B  & $10.0$  & $11.1$ & $+1.1$  & $[-3.6,5.8]$   & $9.3\%\to8.3\%$ \\
FinLLaMa    & $10.0$  & $10.4$ & $+0.3$  & $[-1.9,2.7]$   & $11.2\%\to10.7\%$ \\
Mistral-24B & $-0.3$  & $10.3$ & $\mathbf{+10.6}^{**}$ & $[4.5,17.1]$ & $9.5\%\to8.5\%$ \\
Phi-4       & $-0.8$  & $10.0$ & $+10.8$ & $[1.1,20.4]$  & $10.4\%\to7.8\%$ \\
Qwen2.5-14B & $9.9$   & $10.5$ & $+0.6$  & $[-5.9,7.2]$   & $9.1\%\to8.0\%$ \\
Gemma3-27B  & $-0.8$  & $14.0$ & $\mathbf{+14.8}^{**}$ & $[6.5,23.7]$ & $9.2\%\to7.8\%$ \\
Qwen2.5-32B & $7.8$   & $9.9$  & $+2.1$  & $[-3.9,8.3]$   & $8.9\%\to8.0\%$ \\
\midrule
\textbf{Model-ensemble (joint)} & \textbf{2.7} & \textbf{12.0} & $\mathbf{+9.2}$ & $\mathbf{[4.1,14.6]}$ & $8.8\%\to7.8\%$ \\
\bottomrule
\end{tabular}
\end{table*}

Of the model-ensemble $+9.2$pp/yr effect, $+9.0$ is attributable to the mean-return component and $+0.2$ to
the variance component. The matched improvement is positive in $9/9$ models, and $3/9$ (Qwen2.5-7B,
Mistral-24B, Gemma3-27B) individually survive Holm correction. The joint effect is strongly
significant ($p<0.001$). Realized volatility falls in $9/9$. The model-ensemble effect also remains positive
and significant excluding Qwen2.5-7B ($+5.5$pp/yr, 95\% CI $[1.0,10.1]$, $p=0.015$, same
joint-bootstrap method); Qwen2.5-7B's large contribution does not alone drive the model-ensemble
result. The model-ensemble raw system returns $3.9\%$/yr versus $12.9\%$/yr for the matched system
(Table~\ref{tab:baselines-compact}). Beta regression in place of isotonic gives an almost identical
effect ($+9.4$pp/yr, CI $[4.9,14.3]$), with more models Holm-significant ($5/9$ vs.\ $3/9$).

\paragraph{Case study: Qwen2.5-7B} Qwen exhibits the clearest mismatch: its raw system realizes
$11.7$--$13.8\%$ volatility against the shared $5\%$ calibration target, running at $2.3$--$2.8\times$
its intended risk. Its full contribution is included in the $+9.2$pp/yr headline effect; we treat the
case as an illustrative governance failure of the \emph{raw} system rather than as a caveat on the
matched system's validity. Excluding it entirely, the ensemble effect remains $+5.5$pp/yr
$[1.0,10.1]$, so the result is not driven by one case.

\begin{table}[t]
\caption{Baseline comparison, 9-model-ensemble $\gamma=3$ CER, Sharpe, and realized volatility,
2022--2023 test. Sign-only ignores confidence entirely (equal weight on every active direction).
Adaptive rows share the identical adaptive-volatility controller as Table~\ref{tab:mechanism}'s common
controller row. Raw/matched frozen-scale rows equal Table~\ref{tab:matched} exactly. Sign-only/adaptive
rows use a separate re-fit under the identical protocol.}
\label{tab:baselines-compact}
\footnotesize
\setlength{\tabcolsep}{5pt}
\begin{tabular}{@{}lrrr@{}}
\toprule
System & CER (pp/yr) & Sharpe & Vol. \\
\midrule
\multicolumn{4}{@{}l}{\emph{Frozen-scale controller}} \\
\quad Sign-only & $-0.8$ & $0.05$ & $9.0\%$ \\
\quad Raw & $+2.7$ & $0.44$ & $8.8\%$ \\
\quad \textbf{Matched} & $\mathbf{+12.0}$ & $\mathbf{1.64}$ & $7.8\%$ \\
\midrule
\multicolumn{4}{@{}l}{\emph{Adaptive-volatility controller}} \\
\quad Raw & $+6.6$ & $1.42$ & $4.9\%$ \\
\quad Correctness map & $+8.2$ & $1.86$ & $4.6\%$ \\
\bottomrule
\end{tabular}
\end{table}

Raw confidence adds little beyond sign-only sizing: $-0.8$pp/yr CER for sign-only versus
$+2.7$pp/yr for the LLM's own uncalibrated magnitude (Table~\ref{tab:baselines-compact}). The
gain appears once the correctness map is paired with its own frozen-scale controller ($\kappa$
fitted during calibration and frozen during evaluation; $+12.0$pp/yr), also
reducing maximum drawdown from $-11.7\%$ to $-4.9\%$ and improving daily $5\%$ expected shortfall
from $-1.26\%$ to $-1.03\%$.

We also test an adaptive-volatility controller that re-scales $\kappa$ online in non-overlapping
$21$-session ($\sim$monthly) blocks toward the same $5\%$ target: after each block it applies a
multiplicative log-scale update $\log\kappa_{b+1}=\log\kappa_b-\gamma\,e_b$, where $e_b$ is the prior
block's realized-versus-target annualized-volatility gap, using only pre-block information and
clipping $\kappa$ to $[0.2,5]\times$ the frozen scale (learning rate $\gamma$ fit on 2021). Under this
controller the correctness map still exceeds raw ($+8.2$ vs.\ $+6.6$pp/yr). The matched map thus
adds substantial value under the frozen-scale controller, but only limited and statistically
imprecise incremental value under the adaptive-volatility controller (Section~\ref{sec:mechanism}).

\subsection{Magnitude reallocation, not filtering, drives most of the matched effect}\label{sec:mechanism}

\paragraph{Mechanism check.} Most of the matched effect remains when the raw and correctness-map systems
are forced onto common trade support.

Because $h(u)=\max(0,2u-1)$ embeds a boundary at $u=0.5$, the full rule may benefit from removing
trades as well as reallocating magnitude. We therefore isolate three systems, each with its own fitted
$\kappa$: \emph{gate-only} ($h(u)=\mathbf 1[\hat\mu(u)>0.5]$, equal weight among retained positions,
all magnitude information discarded), \emph{common-support} ($h(u)=\max(0,2\hat\mu(u)-1)$ with zeros
replaced by the smallest positive calibration-split size, preserving every raw active position), and
the full matched system.

\begin{table}[t]
\caption{Mechanism, boundary, and robustness summary, ensembling the nine-model grid unless noted
(Sections~\ref{sec:robust-compact}--\ref{sec:mechanism}).}
\label{tab:mechanism}
\footnotesize
\setlength{\tabcolsep}{4pt}
\begingroup
\renewcommand{\arraystretch}{1.04}
\begin{tabular}{@{}p{0.30\linewidth}p{0.32\linewidth}p{0.31\linewidth}@{}}
\toprule
Diagnostic & Specification & Finding \\
\midrule
Common trade support & Gate-only vs.\ common-support & $+3.9$ vs.\ $+8.5$pp/yr; common-support recovers $93\%$ \\[1pt]
Magnitude compression & Dispersion temp.\ vs.\ Beta vs.\ isotonic & $11\%$ vs.\ $91\%$ vs.\ $100\%$ of isotonic's gain \\[1pt]
Shuffled-label refit & $300$ shuffled-label map--scale refits & Observed $+9.2$pp exceeds $299/300$ draws; excess $+3.6$pp, perm.\ interval $[0.7,7.3]$, $p{=}0.0066$ \\[1pt]
Within-date shuffled refit & $300$ within-date shuffled-label refits & Observed $+9.2$pp exceeds $300/300$ draws; excess $+3.2$pp, $p{=}0.0033$ \\[1pt]
Adaptive controller & Correctness-adaptive $-$ raw-adaptive & $+1.6$pp/yr, CI $[-0.8,4.1]$, $p{=}0.195$ \\[1pt]
Controller interaction & Frozen matched effect $-$ adaptive matched effect & $+7.78$pp/yr, CI $[2.01,15.25]$, $p{<}0.001$ \\[1pt]
Adaptive label-free baseline & Matched system vs.\ label-free adaptive-volatility-only & $+5.4$pp/yr, CI $[-0.1,10.9]$, $p{=}0.056$ \\
\bottomrule
\end{tabular}
\endgroup
\end{table}

The matched effect remains positive across alternative raw sizing maps and seven model-family
compositions, while direct-payoff estimation exceeds correctness calibration in $10$ of $18$
conditions.

\paragraph{Robustness.} Table~\ref{tab:mechanism} tests three alternative explanations. Forcing
common trade support leaves most of the gain intact, so magnitude reallocation rather than filtering
drives the result; dispersion matching alone recovers little of isotonic's gain; and although joint
map--scale refitting has a positive shuffled-label baseline, the observed effect exceeds both
unrestricted and within-date shuffled-label refits. Temporal-shift and wrong-ticker placebos likewise
eliminate the effect, and net exposure and market beta change little across books. A
Newey--West regression of the matched-minus-raw spread on tech (non-tech minus tech),
momentum, and volatility terciles leaves an annualized intercept of $+8.6$pp/yr
$[3.2,14.1]$, unexplained by these exposures.

\paragraph{Deployment boundary (adaptive controller).} Against a label-free adaptive-volatility-only
baseline (the raw map under the adaptive controller defined above), the
matched system is directionally better in $9/9$ models; the estimate is positive but statistically
imprecise (Table~\ref{tab:mechanism}). A paired test of per-model effects across the two controllers
confirms that the matched-system effect is significantly larger under frozen-scale control
(interaction: $+7.78$pp/yr, 95\% CI $[2.01, 15.25]$, $p<0.001$).

\subsection{Temporal transfer depends on fitting regime}\label{sec:walkforward}

\paragraph{Deployment boundary (annual transfer).} Annual refitting yields smaller net gains because positive
map--scale interaction is offset by adverse scale transfer.

\begin{figure*}[t]
\centering
\includegraphics[width=0.70\linewidth]{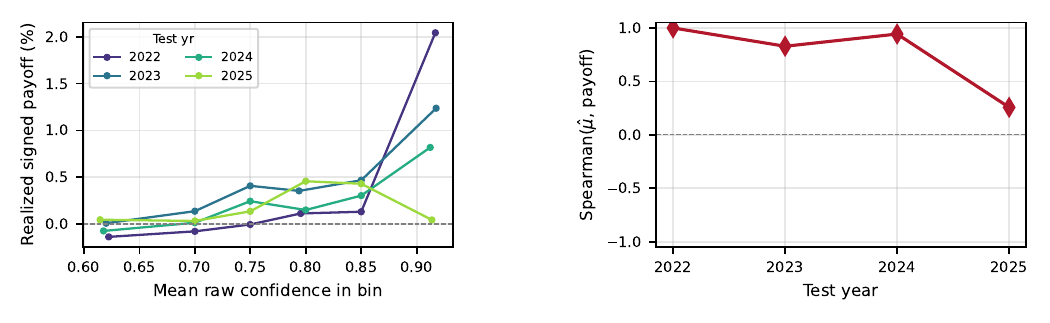}\\[4pt]
\includegraphics[width=0.78\linewidth]{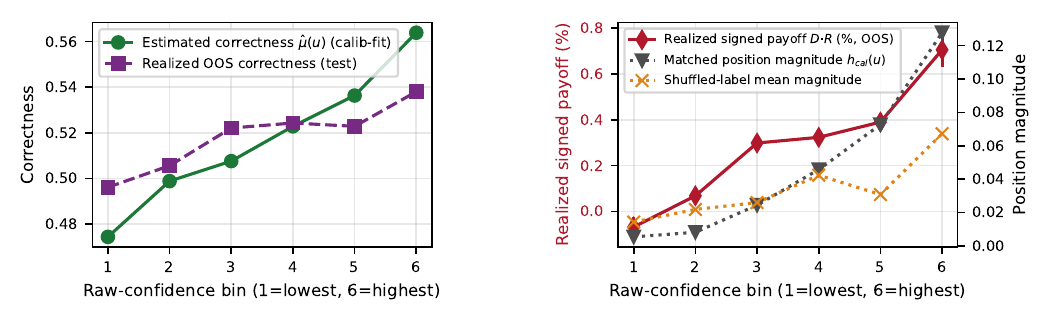}
\caption{Correctness--payoff diagnostics. \textbf{Top:} alignment across the four annual walk-forward
folds (five-model roster, test years 2022--2025); left, realized signed payoff $D\cdot R$ by
raw-confidence bin, one line per year; right, cross-bin Spearman correlation between each fold's fitted
correctness map $\hat\mu(u)$ and realized signed payoff, the 2025 fold breaking down because its
highest-confidence bin's payoff no longer exceeds lower bins'. \textbf{Bottom:} bridge across the 18
cardinal model-condition cells (test period); calibration-fitted correctness tracks test-period
correctness (left), while realized signed payoff generally rises with matched-system position
magnitude but not shuffled-label magnitude (right).}
\label{fig:bridge-wf}
\label{fig:bridge}
\end{figure*}

We evaluate temporal transfer under two protocols. Under the annual walk-forward protocol of
Section~\ref{sec:data} (refit on year $t$, evaluate frozen on $t{+}1$; five models, named and
anonymized), all three calibrators are directionally positive but none survives family-wise Holm
correction ($p=0.144$, $0.058$, $0.123$ for isotonic, Beta, Platt; Beta most precise;
Table~\ref{tab:walkforward}). Transfer is therefore positive but attenuated and estimator-dependent
rather than regime-invariant. The exploratory fixed-window protocol instead refits once on 2024H1 and produces larger isotonic
effects in 2024H2 and 2025 ($+21.3$ and $+15.7$pp/yr; nominal, uncorrected), reinforcing sensitivity
to calibration-window choice. The attenuation coincides with a deterioration in the confidence
signal: pooled Murphy resolution (RES/UNC) for the five-model verbalized channel falls from
$0.17$--$0.19\%$ in 2022--2023 to $0.04$--$0.08\%$ in 2024--2025, consistent with contemporaneous
evidence of weakening LLM return predictability~\citep{lopezlira2023}. Correctness--payoff alignment
likewise falls sharply (Figure~\ref{fig:bridge-wf}, top: $1.00$/$0.83$/$0.94$ in 2022--2024, $0.26$ in
2025). These diagnostics do not explain the fold-level $\Delta$CER pattern, however: the weakest fold
($2023$, $-0.70$pp/yr) has the best resolution and second-best alignment, coinciding instead with an
unusually strong underlying-universe return ($+28.7\%$ annualized) under net-long books. We therefore
treat resolution and alignment as descriptive rather than causal explanations of transfer.

\begin{table}[t]
\caption{Annual walk-forward transfer, model-ensemble $\Delta$CER ($\gamma=3$, pp/yr) across four disjoint
annual folds (2022--2025), five models, named+anon. Ensemble $\Delta$CER/CI is a pooled-return
bootstrap (concatenated across folds, block-respecting); the Stouffer test below separately checks
replication across folds (two-sided, $n$-weighted $Z$, Holm-corrected across the three calibrators).
Calibrators are listed in declaration order, not sorted by performance.}
\label{tab:walkforward}
\footnotesize
\setlength{\tabcolsep}{3pt}
\begin{tabular}{@{}lrrrr@{}}
\toprule
Calibrator & Ensemble $\Delta$CER [95\% CI] & $Z$ & Comb.\ $p$ & Holm $p$ \\
\midrule
Isotonic & $+1.17\ [-0.60,3.06]$ & $1.46$ & $0.144$ & $0.144$ \\
Beta     & $+1.50\ [0.30,2.70]$  & $2.34$ & $0.019$ & $0.058$ \\
Platt    & $+0.67\ [-0.26,1.61]$ & $1.87$ & $0.061$ & $0.123$ \\
\bottomrule
\end{tabular}
\end{table}

\paragraph{Decomposing the walk-forward effect.} Under isotonic calibration, the pooled annual
effect decomposes into $-4.0$pp/yr from map replacement, $-8.1$pp/yr from scale transfer, and
$+13.3$pp/yr from their interaction (eq.~\eqref{eq:decomp}, fold-respecting block bootstrap): the
small net gain reflects offsetting components rather than the disappearance of map--scale
dependence, and the interaction is positive in every fold.

\section{Limitations and Conclusion}\label{sec:limits}\label{sec:conclusion}

\paragraph{Signal quality.} The base classifier is weak ($50.3$--$51.8\%$ accuracy vs.\ a
$53.8\%$ base rate), so we make no market-beating claim and treat Table~\ref{tab:baselines-compact}'s
sign-only book as the relevant floor comparator. Correctness is used only as the lower-variance
payoff proxy described in Section~\ref{sec:cardinal}; pooled diagnostics across all $18$ configurations
(Figure~\ref{fig:bridge}, bottom) show fitted correctness and signed payoff generally rising together, but do
not establish this relationship within every model or masking condition. Because joint refitting has a
positive shuffled-label baseline (Section~\ref{sec:mechanism}), the $+9.2$pp/yr effect is best read
as evidence for the fitted system rather than a purely information-driven effect.

\paragraph{Temporal validity.} All nine models were released after the primary $2022$--$2023$
evaluation period, so pretraining exposure to those market outcomes cannot be completely ruled out;
only the 2025 observations are verified to postdate the release dates of the two primary models
(Qwen2.5-7B, Gemma-2-9B), and training-cutoff audit status for the other seven is not fully
documented. A named-vs-anonymized comparison shows no gap ($+9.1$ vs.\ $+9.4$pp/yr, pooled): if the
gain were ticker-identity memorization, masking names/tickers should attenuate it; it does not. More
fundamentally, the paper's central claim does not require the
$2022$--$2023$ trading gain to be contamination-free: the map--scale coupling result
(Section~\ref{sec:robust-compact}) follows from how a fixed rule responds to a confidence-map change
(Section~\ref{sec:theory}), not any one period's return sign or size. Contamination could inflate the
headline effect's magnitude but not the analytical non-portability result. We cannot rule out contamination that does not require
entity recognition (e.g., recalled macro/sector narratives), and annual walk-forward gains are smaller
and do not survive family-wise correction (Section~\ref{sec:walkforward}).

\paragraph{Generalizability.} Only $3$ of $9$ per-model effects individually survive Holm, so the
result concerns matched-system levels and map--scale compatibility, not per-model significance.
Pre-declared partitions concentrate essentially all of the primary-period gain in $2022$
($+16.2$pp/yr, $p<0.001$) and non-technology tickers ($+13.9$pp/yr, $p=0.002$, $34/61$ names),
with $2023$ and technology names both insignificant---a real but partial tilt, not a full
explanation of the effect (factor regression, Section~\ref{sec:mechanism}). The five-session hold
was fixed a priori as the primary weekly cadence; the twelve-horizon sensitivity sweep
(Section~\ref{sec:data-splits}) is descriptive and not covered by the Holm/Romano--Wolf corrections,
and results are stable under $1$--$3\%$ annual short-borrow costs (model-ensemble gains
$+9.5$--$9.8$pp/yr).

\paragraph{Conclusion.} Monotone recalibration does not affect coverage-matched gates because it
preserves score ordering. Position sizing is different: changing the confidence map changes portfolio
exposures and may invalidate a scale fitted to the original score distribution. In our data,
matching the map and scale materially improves the frozen-scale system; the benefit attenuates
under adaptive risk control and later walk-forward evaluation. Confidence maps should
therefore be monitored and revalidated as part of the full portfolio-construction system.

\begingroup
\footnotesize
\bibliographystyle{ACM-Reference-Format}
\bibliography{refs}

@article{chow1970,
  author = {Chow, C. K.},
  title = {On Optimum Recognition Error and Reject Tradeoff},
  journal = {IEEE Transactions on Information Theory},
  volume = {16},
  number = {1},
  pages = {41--46},
  year = {1970}
}

@inproceedings{geifman2017,
  author = {Geifman, Yonatan and El-Yaniv, Ran},
  title = {Selective Classification for Deep Neural Networks},
  booktitle = {Advances in Neural Information Processing Systems},
  series = {NeurIPS},
  pages = {4878--4887},
  year = {2017}
}

@article{degroot1983,
  author = {DeGroot, Morris H. and Fienberg, Stephen E.},
  title = {The Comparison and Evaluation of Forecasters},
  journal = {The Statistician (Journal of the Royal Statistical Society, Series D)},
  volume = {32},
  number = {1--2},
  pages = {12--22},
  year = {1983}
}

@article{murphy1973,
  author = {Murphy, Allan H.},
  title = {A New Vector Partition of the Probability Score},
  journal = {Journal of Applied Meteorology},
  volume = {12},
  number = {4},
  pages = {595--600},
  year = {1973}
}

@article{brier1950,
  author = {Brier, Glenn W.},
  title = {Verification of Forecasts Expressed in Terms of Probability},
  journal = {Monthly Weather Review},
  volume = {78},
  number = {1},
  pages = {1--3},
  year = {1950}
}

@inproceedings{guo2017,
  author = {Guo, Chuan and Pleiss, Geoff and Sun, Yu and Weinberger, Kilian Q.},
  title = {On Calibration of Modern Neural Networks},
  booktitle = {International Conference on Machine Learning},
  series = {ICML},
  pages = {1321--1330},
  year = {2017}
}

@inproceedings{naeini2015,
  author = {Naeini, Mahdi Pakdaman and Cooper, Gregory F. and Hauskrecht, Milos},
  title = {Obtaining Well Calibrated Probabilities Using {Bayesian} Binning},
  booktitle = {AAAI Conference on Artificial Intelligence},
  pages = {2901--2907},
  year = {2015}
}

@article{lopezlira2023,
  author  = {Lopez-Lira, Alejandro and Tang, Yuehua},
  title   = {Can {ChatGPT} Forecast Stock Price Movements? {R}eturn Predictability and Large Language Models},
  journal = {Journal of Financial Economics},
  volume  = {184},
  pages   = {104335},
  year    = {2026}
}

@inproceedings{zhao2021decision,
  author = {Zhao, Shengjia and Kim, Michael P. and Sahoo, Roshni and Ma, Tengyu and Ermon, Stefano},
  title = {Calibrating Predictions to Decisions: A Novel Approach to Multi-Class Calibration},
  booktitle = {Advances in Neural Information Processing Systems},
  series = {NeurIPS},
  year = {2021}
}

@misc{bas2026,
  author = {Wu, Sean and Gustafsson, Fredrik K. and Phillips, Edward and Gao, Boyan and Thakur, Anshul and Clifton, David A.},
  title = {{BAS}: A Decision-Theoretic Approach to Evaluating Large Language Model Confidence},
  howpublished = {arXiv:2604.03216},
  year = {2026}
}

@misc{glasserman2023,
  author = {Glasserman, Paul and Lin, Caden},
  title = {Assessing Look-Ahead Bias in Stock Return Predictions Generated by {GPT} Sentiment Analysis},
  howpublished = {arXiv:2309.17322},
  year = {2023}
}

@misc{chen2024,
  author = {Chen, Shuaiyu and Green, T. Clifton and Gulen, Huseyin and Zhou, Dexin},
  title = {What Does {ChatGPT} Make of Historical Stock Returns? {E}xtrapolation and Miscalibration in {LLM} Stock Return Forecasts},
  howpublished = {arXiv:2409.11540},
  year = {2024}
}

@article{romano2005,
  author = {Romano, Joseph P. and Wolf, Michael},
  title = {Stepwise Multiple Testing as Formalized Data Snooping},
  journal = {Econometrica},
  volume = {73},
  number = {4},
  pages = {1237--1282},
  year = {2005}
}

@misc{kadavath2022,
  author = {Kadavath, Saurav and others},
  title = {Language Models (Mostly) Know What They Know},
  howpublished = {arXiv:2207.05221},
  year = {2022}
}

@misc{wang2022self,
  author = {Wang, Xuezhi and others},
  title = {Self-Consistency Improves Chain of Thought Reasoning in Language Models},
  howpublished = {arXiv:2203.11171},
  year = {2022}
}

@inproceedings{tian2023,
  author = {Tian, Katherine and others},
  title = {Just Ask for Calibration: Strategies for Eliciting Calibrated Confidence Scores from Language Models Fine-Tuned with Human Feedback},
  booktitle = {Empirical Methods in Natural Language Processing},
  series = {EMNLP},
  address = {Singapore},
  pages = {5433--5442},
  year = {2023}
}

@inproceedings{xiong2024,
  author = {Xiong, Miao and others},
  title = {Can {LLMs} Express Their Uncertainty? {A}n Empirical Evaluation of Confidence Elicitation in {LLMs}},
  booktitle = {International Conference on Learning Representations},
  series = {ICLR},
  year = {2024}
}

@inproceedings{joy2023adaptive,
  author = {Joy, Tom and Pinto, Francesco and Lim, Ser-Nam and Torr, Philip H. S. and Dokania, Puneet K.},
  title = {Sample-Dependent Adaptive Temperature Scaling for Improved Calibration},
  booktitle = {Proceedings of the AAAI Conference on Artificial Intelligence},
  pages = {14919--14926},
  year = {2023}
}

@inproceedings{rabanser2025selective,
  author = {Rabanser, Stephan and Papernot, Nicolas},
  title = {What Does It Take to Build a Performant Selective Classifier?},
  booktitle = {Advances in Neural Information Processing Systems},
  year = {2025}
}

@article{markowitz1952,
  author = {Markowitz, Harry},
  title = {Portfolio Selection},
  journal = {The Journal of Finance},
  volume = {7},
  number = {1},
  pages = {77--91},
  year = {1952}
}

@article{kelly1956,
  author = {Kelly, J. L.},
  title = {A New Interpretation of Information Rate},
  journal = {Bell System Technical Journal},
  volume = {35},
  number = {4},
  pages = {917--926},
  year = {1956}
}

@book{grinoldkahn2000,
  author = {Grinold, Richard C. and Kahn, Ronald N.},
  title = {Active Portfolio Management: A Quantitative Approach for Producing Superior Returns and Controlling Risk},
  edition = {2nd},
  publisher = {McGraw-Hill},
  year = {2000}
}

@article{moreiramuir2017,
  author = {Moreira, Alan and Muir, Tyler},
  title = {Volatility-Managed Portfolios},
  journal = {The Journal of Finance},
  volume = {72},
  number = {4},
  pages = {1611--1644},
  year = {2017}
}

@article{demiguel2009,
  author = {DeMiguel, Victor and Garlappi, Lorenzo and Uppal, Raman},
  title = {Optimal Versus Naive Diversification: How Inefficient Is the 1/N Portfolio Strategy?},
  journal = {The Review of Financial Studies},
  volume = {22},
  number = {5},
  pages = {1915--1953},
  year = {2009}
}

@misc{iacovides2025findpo,
  author = {Iacovides, Giorgos and Zhou, Wuyang and Mandic, Danilo},
  title = {FinDPO: Financial Sentiment Analysis for Algorithmic Trading through Preference Optimization of LLMs},
  howpublished = {arXiv:2507.18417},
  year = {2025}
}

@misc{vamvourellis2025reasoning,
  author = {Vamvourellis, Dimitris and Mehta, Dhagash},
  title = {Reasoning or Overthinking: Evaluating Large Language Models on Financial Sentiment Analysis},
  howpublished = {arXiv:2506.04574},
  year = {2025}
}

@misc{dimino2025positional,
  author = {Dimino, Fabrizio and Saxena, Krati and Sarmah, Bhaskarjit and Pasquali, Stefano},
  title = {Tracing Positional Bias in Financial Decision-Making: Mechanistic Insights from {Qwen2.5}},
  howpublished = {arXiv:2508.18427},
  year = {2025}
}

@misc{lee2025yourai,
  author = {Lee, Hoyoung and Seo, Junhyuk and Park, Suhwan and Lee, Junhyeong and Ahn, Wonbin and Choi, Chanyeol and Lopez-Lira, Alejandro and Lee, Yongjae},
  title = {Your {AI}, Not Your View: The Bias of {LLM}s in Investment Analysis},
  howpublished = {arXiv:2507.20957},
  year = {2025}
}
\endgroup

\end{document}